\documentclass[12pt]{iopart}

\renewcommand{\a}{\alpha}
\renewcommand{\b}{\beta}
\renewcommand{\c}{\gamma}
\renewcommand{\d}{\delta}

\newcommand{\g}{\gamma}

\newcommand{\half}{\frac{1}{2}}
\newcommand{\ad}{\dot\alpha}

\newcommand{\p}{\partial}

\newcommand{\s}{\sigma}

\renewcommand{\t}{\theta}

\newcommand{\be}{\begin{equation}}
\newcommand{\eq}{\end{equation}}

\def\ad{\dot\alpha}
\def\half{\frac{1}{2}}

\def\beqa{\begin{eqnarray}}
\def\eeqa{\end{eqnarray}}

\begin{document}
\jl{6}

\begin{center}
\hfill {IFT.P.078/99}\\
\end{center}

\title{ Quantization of the Superstring in Ramond-Ramond Backgrounds}

\author{Nathan Berkovits}

\address{
Instituto de F\'\i sica Te\'orica, Universidade Estadual
Paulista\\
Rua Pamplona 145, 01405-900, S\~ao Paulo, SP, Brasil}

\author{Talk presented at Strings `99 (Potsdam, Germany)}

\begin{abstract}

Sigma model actions are constructed for the Type II
superstring compactified to
four and six dimensional curved backgrounds which can
contain non-vanishing Ramond-Ramond fields.
These actions are N=2 worldsheet superconformally invariant and can
be covariantly quantized preserving manifest spacetime supersymmetry.
They are constructed using a hybrid version of superstring variables
which combines features of the Ramond-Neveu-Schwarz and Green-Schwarz
formalisms. For the $AdS_2\times S^2$ and $AdS_3\times S^3$ backgrounds,
these actions
differ from the classical Greeen-Schwarz actions by a crucial
kinetic term for the fermions.
Parts of this work have been done in collaborations with
M. Bershadsky, T. Hauer, W. Siegel, C. Vafa, E. Witten, S. Zhukov and
B. Zwiebach.

\end{abstract}

\maketitle

\section{Introduction}

Quantization of the superstring in Ramond-Ramond backgrounds
is currently of great interest because of the AdS-CFT conjectures.
Although Ramond vertex operators are well
understood
using the results of Friedan, Martinec and Shenker \cite{FMS} for the
Ramond-Neveu-Schwarz (RNS) formalism, these vertex operators
differ from  Neveu-Schwarz vertex operators in breaking worldsheet
supersymmetry and mixing matter and ghost variables.
Without worldsheet supersymmetry
as a guiding principle, it is difficult to guess the correct
generalization of these vertex operators for constructing
the superstring action
in Ramond-Ramond backgrounds.

It has been proposed in \cite{contact} that spacetime supersymmetry may
replace worldsheet supersymmetry as a guiding principle
for constructing the action,
e.g. in
determining the `contact terms' which are required to eliminate
unphysical divergences\cite{seiberg}. 
However, spacetime supersymmetry
is hard to verify in the RNS formalism since the worldsheet action is
only expected to be spacetime-supersymmetric when
the background fields are on-shell. This can be seen from the fact
that spacetime supersymmetry transformations change the `picture' of a
vertex operator, but picture-changing can only be defined when
the vertex operator is physical.

An alternative approach is to use the Green-Schwarz (GS) formalism
for the superstring where spacetime supersymmetry is manifest.
Although it is easy to construct a classical GS action for the
superstring in a curved superspace background including
Ramond-Ramond fields, the existence of fermionic
second-class constraints makes it difficult 
to quantize this action unless the background
admits a light-cone gauge choice. Furthermore, the lack of
a Fradkin-Tseytlin term coupling the dilaton to worldsheet curvature
makes it doubtful that the classical GS action is correct in
arbitrary backgrounds.

The fermionic second-class constraints of the GS formalism 
are 
$$d_\a = p_\a - i \s_{\a\b}^m \t^\a \p x_m + ... =0$$
where $(x_m,\t^\a)$ are the usual superspace variables, $p_\a$ is
the conjugate momentum for $\t^\a$, $\s_{\a\b}^m$ are the
Pauli matrices, and $...$ refers to terms which are higher order in
$\t^\a$.
Siegel has suggested in \cite{siegelclass} a
modification of the GS formalism in which the above
second class constraints are replaced by a suitable set of
first-class constraints constructed from $d_\a$. In this modified GS formalism,
$p_\a$ is treated as an independent worldsheet variable.
Although he was unable to find an appropriate set of first-class constraints,
he argued on general grounds that the massless open superstring vertex
operator should have the form
$$
\label{openvertex}
V = \int dz [ \p Y^M ~A_M(x,\t) + d_\a ~W^\a(x,\t) ]
$$
where $Y^M=(x^m,\t^\a)$, $A_M$ are the superspace gauge fields,
and $W^\a$ are the superspace field-strengths.

Over the last five years, a third approach has been developed
which combines advantages of the RNS and GS approaches and
has therefore been named the `hybrid' approach. Using a field redefinition
which preserves free-field OPE's,
one can map the worldsheet matter and ghost RNS variables into
a set of superspace variables (including conjugate $p_\a$ variables)
plus an internal sector. In fact, part of this map was
already suggested in \cite{FMS} where it was 
noticed that since the spacetime-supersymmetry generator in
the $-\half$ picture is given by $q_\a =\int dz
 e^{-\frac{1}{2}\phi}\Sigma_\a$, it is natural to define
$$\t^\a =
 e^{\frac{1}{2}\phi}\Sigma^\a, \quad p_\a = 
 e^{-\frac{1}{2}\phi}\Sigma_\a, $$
where $\phi$ comes from fermionizing the bosonic ghosts
as $\beta =\p\xi e^{-\phi}$,
$\gamma=\eta e^\phi$ and $\Sigma^\a$ is the spin field of conformal
weight $5/8$ constructed from
the $\psi^m$ variables.

Unfortunately, these $\t^\a$ variables are not all free fields if
$\a =1$ to 16 as can be seen from the OPE
$$\t^\a (y) \t^\b( z) \to (y-z)^{-1} \s^{\a\b}_m  e^{\phi}\psi^m.$$ 
However, it is possible to choose a subset of these
variables which are free fields. The subset that is most convenient depends
on which subgroup of $SO(9,1)$ Lorentz invariance one wants to remain manifest. 
For example, using $U(5)$ notation, the subset 
[$\t^{+++++}, \t^{--+++}$] are free fields which leave
manifest the subgroup $SO(3,1)\times U(3)$. This choice is 
most convenient for describing compactifications of the superstring
to four dimensions on a Calabi-Yau three-fold \cite{four}\cite{review}\cite
{siegel}\cite{zw}. 
Another choice is the subset
[$\t^{+++++}, \t^{--+++}, \t^{+--++},\t^{-+-++}$] which leaves manifest
the subgroup $SO(5,1)\times U(2)$.
This choice 
is most convenient for describing compactifications of the superstring
to six dimensions on a Calabi-Yau two-fold \cite
{vafa}
\cite{witten} \cite{six}.
A third choice is
the subset 
[$\t^{+----}, \t^{-+---}, \t^{--+--},\t^{---+-}, \t^{----+}$] 
which leaves manifest
a $U(5)$ subgroup of the Wick-rotated
$SO(10)$ \cite{ufive}. 

In section 2, the hybrid formalism for compactification to four dimensions
is used to construct the superstring action
in an $AdS_2\times S^2$
background with Ramond-Ramond flux. 
In section 3, the hybrid formalism for compactification
to six dimensions is used to construct the superstring action
in an $AdS_3\times S^3$ background with Ramond-Ramond flux. 
Work is in progress on using the $U(5)$ version of the hybrid
formalism to construct the superstring action
in an $AdS_5\times S^5$ background with Ramond-Ramond flux.

\section{Hybrid formalism for compactification to four dimensions}

After defining $\t^\a = [ e^{\frac{1}{2}\phi}\Sigma^{+++++},
e^{\frac{1}{2}\phi}\Sigma^{--+++}]$ and
$p_\a = [ e^{-\frac{1}{2}\phi}\Sigma^{-----},
e^{-\frac{1}{2}\phi}\Sigma^{++---}]$ to be fundamental variables for
$\a=1$ to 2,  
one still has to define the remaining set of hybrid variables. It is
convenient to define 
$$\hat\t^{\dot \a}= [c \xi e^{-\frac{3}{2}\phi} 
\Sigma^{+----},
 c \xi e^{-\frac{3}{2}\phi} 
\Sigma^{-+---}],$$
$$\hat p_{\dot \a}= [b \eta e^{\frac{3}{2}\phi} 
\Sigma^{-++++},
 b \eta e^{\frac{3}{2}\phi} 
\Sigma^{+-+++}]$$
to be the complex conjugates of $\t^\a$ and $p_\a$. As discussed
in \cite{herm}, this definition of complex conjugation flips the
picture so $\hat q^{\dot\a}$ is naturally defined in the $+\half$
picture as 
$$\hat q^{\dot\a}= \int dz [ b\eta e^{\frac{3}{2}\phi} \Sigma^{\dot\a}
+e^{\frac{1}{2}\phi} \p x^m \s_m^{\a\dot\a} \Sigma^{\a}].$$
With this choice of picture for the four-dimensional supersymmetry
generators, $\{q^\a, \hat q^{\dot\a}\}= \int dz \p x^m \s_m^{\a\ad}$
as desired for the supersymmetry algebra.

For compactifications of the superstring which
preserve four-dimensional supersymmetry, the six-dimensional
compactification manifold
can be represented by a $c=9$  N=2 superconformal field theory whose
N=2 generators will be denoted $[T_C, G^+_C, G^-_C, J_C]$.
Up to some minor
modifications which will be described below, the compactification
manifold in the hybrid formalism is represented by the same $c=9$
N=2 superconformal field theory.
For the compactification-dependent worldsheet fields to have no singular OPE's
with $\t^\a$ and $\hat\t^{\dot\a}$, they need to be twisted
by a factor $e^{N\kappa }$ where $\p\kappa=\p\phi + \eta\xi$
and $N$ is the charge of the
field with respect to $J_C$.
For example, for compactification on a six-torus, the worldsheet fields
$\Gamma^j=\psi^{j+3} + i\psi^{j+6}$ for $j=1$ to 3 are redefined to 
$$\Gamma^j \to e^{\kappa}\Gamma^j, \quad
\bar\Gamma_j \to e^{-\kappa}\bar\Gamma_j$$
which have no singular OPE's with $\t^\a$ and 
$\hat\t^{\dot\a}$. One also needs to redefine the $c=9$ N=2 generators to be
$$[T_C, G^+_C, G^-_C, J_C] \to [T_C +\frac{3}{2}(\p\kappa)^2-\p\kappa J_C,
~e^\kappa G^+_C,~ e^{-\kappa} G^-_C, ~J_C +3\p\kappa]$$
which still generates a $c=9$ N=2 algebra.

There is one chiral boson $\rho$ defined by
$$\p\rho= -3\p\phi +cb +2\xi\eta - J_{C}$$
which has no singular OPE's with the other hybrid variables. This is the last
free hybrid variable since the compactification-independent hybrid variables
$[x^m, \t^\a,\hat\t^{\dot\a}, p_\a, \hat p_{\dot\a}, \rho]$
contain the same degrees of freedom as the compactification-independent
RNS variables $[x^m, \psi^m, b, c, \xi, \eta, \phi]$ where $m=0$ to 3.
So any vertex operator constructed out of RNS variables can
be rewritten in terms of hybrid variables and vice versa.

One now needs to define the physical state conditions in
terms of the hybrid variables. In terms of the RNS variables,
the physical state conditions for a vertex operator $V$ are 
$Q V =q_{ghost} V =0$ where $Q_{BRST}=\int dz j_{BRST}$
is the BRST charge and $q_{ghost}=\int dz j_{ghost}$ is the ghost current. 
At least for Neveu-Schwarz integrated vertex operators in the zero picture,
these conditions can be strengthened to require that $V$ has no poles
with $j_{BRST}$, with $j_{ghost}$, with the $b$ ghost, and with the
stress tensor $T$. 
After adding a total derivative to $j_{BRST}$ and
defining $j_{ghost} = bc +\xi\eta$ (which is equivalent
to the standard definition $j_{ghost}= bc +\p\phi$ in the zero picture), 
one can show that $[T, j_{BRST}, b, j_{ghost}]$ form a twisted N=2 algebra.
So the above definition of physical state conditions implies that $V$ is
a U(1)-neutral primary field with respect to these N=2 generators.
Note that $V$ has no poles with $\eta$ since $\{ b, e^{\int^z j_{ghost}} \}=
\{ b, c\eta \}=\eta$. So it is independent of the $\xi$ zero mode as
desired for RNS physical vertex operators.

After performing a similarity transformation on the hybrid variables of
the form $Y \to e^R Y e^{-R}$ with $R= -\int dz [\t^\a \hat\t^{\dot\a}
\s^m_{\a\dot\a} \p x_m + e^{-\rho}\t^\a\t_\a G^-_{C}]$,
one finds that these twisted N=2 generators are mapped to
$$T =\frac{1}{2} 
\p x^m \p x_m + p_\a \p\theta^\a + \hat p_{\dot \a}\hat\t^{\dot\a}
+ \frac{1}{2}\p\rho\p\rho + \frac{1}{2}\p^2\rho + T_C ,$$
$$G^+ = j_{BRST} = d^\a d_\a e^\rho + G^+_C,$$
$$G^- = b  = \hat d^{\dot\a} \hat d_{\dot \a} e^{-\rho} + G^-_C,$$
$$J = j_{ghost} = \p\rho + J_C,$$
where 
$d_\a = p_\a +\frac{i}{2}\s^m_{\a\dot\a}\hat\t^{\dot \a}
\p x_m -\half(\hat\t)^2\p\t_\a
+{1\over 4}\t_\a \p (\hat\t)^2$ and 
$\hat d_{\dot \a}=\hat p_{\dot \a}
-\frac{i}{2}\s^m_{\a\dot\a}\t^\a\p x_m -\half(\t)^2\p\hat\t_{\dot\a}
+{1\over 4}\hat\t_{\dot \a} \p (\t)^2$
are supersymmetric combinations of the four-dimensional
superspace variables. 
Note that the N=2 generators factorize into a set of four-dimensional
N=2 generators and compactification-dependent N=2 generators. These
N=2 generators
provide first-class constraints for $d_\a$ and $\hat d_{\dot \a}$  that
can replace the second-class constraints of the four-dimensional 
GS superstring.

For the open superstring, the massless compactification-independent
states are described by the gluon and gluino of four-dimensional
super-Yang-Mills.
As shown in \cite{four}, the integrated vertex operator for these states
is given by
$$V = \int dz [ \p Y^M ~A_M + d_\a ~W^\a 
+\hat d_{\dot\a}\bar W^{\dot\a}]
$$
exactly as proposed in \cite{siegelclass}
where $Y^M = (x^m,\t^\a,\hat\t^{\dot\a})$,
$W^\a$ and $\bar W^{\dot\a}$ are the chiral and anti-chiral superspace
field-strengths,
and $A_M$ are the superspace gauge fields. One can
check that in Wess-Zumino gauge, this
vertex operator is mapped to the standard Neveu-Schwarz and Ramond
vertex operators for the gluon and gluino where the Weyl gluino
vertex operator
is in the $-\frac{1}{2}$ picture, the gluon vertex operator is in the
0 picture, and the anti-Weyl gluino vertex operator 
is in
the $+\frac{1}{2}$ picture.

Since the closed superstring vertex operator is the holomorphic square
of the open superstring vertex operator, one obtains the following integrated
vertex operator for the massless compactification-independent states
of the Type II superstring \cite{siegel}:
$$S = \int d^2 z  [ \p Y^M \bar\p Y^N ~(G_{MN} +B_{MN})
 + d_a \bar\p Y^N E^a_N + 
\p Y^M \bar d_{\bar a} E_M^{\bar a} + d_a \bar d_{\bar b} F^{a \bar b}]$$
where $Y^M=
(x^m,\t^\a,\hat\t^{\dot\a}, \bar\t^{\bar\a},\bar{\hat\t}{}^{\bar{\dot\a}})$,
$\bar\t^{\bar\a}$ and 
$\bar{\hat\t}{}^{\bar{\dot\a}}$ are the right-moving analogs of the left-moving
$\t^\a$ and $\hat\t^{\dot\a}$, $a=(\a,\dot\a)$ and 
$\bar a=(\bar\a,\bar{\dot\a})$,
and the lowest components of the superfields $G_{mn}, B_{mn},
E_m^{a}, E_m^{\bar a}, F^{a\bar b}$ are the graviton, the anti-symmetric
tensor, the two gravitini, and the Ramond-Ramond bispinor field-strength.

To get the superstring action in a curved background, one simply
interprets the superfields [$G_{MN},B_{MN},E_M^a,E_M^{\bar a},F^{ a\bar b}$]
appearing in $S$ as background superfields. This action is manifestly
super-reparameterization invariant and is expected to be
invariant under N=2 worldsheet superconformal transformations when
the background fields satisfy the appropriate equations of motion.
Note that the first term in $S$ is the usual GS action in a curved
background, but the other terms in $S$ proportional to
$d_a$ and $\bar d_{\bar a}$ are required for quantization since
they provide kinetic terms for the fermions.

Of course, the complete action also contains a contribution from the
compactification-dependent fields but, at string tree-level, one can
consistently choose the four-dimensional background superfields to be
compactification-independent.
So the compactification-dependent contribution can be chosen to
be the same as in the flat four-dimensional case.
One also needs an action for the chiral boson $\rho$, but this
term similarly decouples from the four-dimensional background.
Finally, one needs to add a Fradkin-Tseytlin term to the action to
couple the dilaton to worldsheet curvature. As discussed in
\cite{siegel}, this term can be constructed using chiral and
twisted-chiral spacetime and worldsheet superfields. Although
it has not yet been checked for the Type II superstring
that N=2 worldsheet superconformal invariance at one-loop sigma model
implies the expected superspace equations of motion  for the background
superfields, this has been checked for an analogous sigma model
action for the heterotic superstring \cite{sken}.

To obtain the action for the superstring in an $AdS_2\times S^2$
background 
with constant Ramond-Ramond flux $F^{a\bar b} =  \delta^{a\bar b}$,
one simply plugs the appropriate values for the background fields into the
action. Because $d_a$ and $\bar d_{\bar a}$ are auxiliary fields in the
presence of constant Ramond-Ramond flux, they can be integrated out
to produce the action
$$
S_{AdS} = \int d^2 z 
[ \eta_{cd} J_z^c J^d_{\bar z} + \frac{1}{2} \delta_{ a\bar a}
(J_z^a J_{\bar z}^{\bar a} - J_{\bar z}^a J_z^{\bar a}) 
$$
$$
+ \delta_{ a\bar a}(J_z^a J_{\bar z}^{\bar a} + 
J_{\bar z}^a J_z^{\bar a})]
$$
where $(J^a, J^{\bar a})$ and $J^c$ are the eight fermionic currents 
and four bosonic currents $(g^{-1} d g)^A$
constructed from a coset supergroup $g$ taking values in $PSU(1,1|2)/U(1)\times
U(1)$ \cite{zw}. 
The first line of this action comes from the $\p Y^M \bar\p Y^N 
(G_{MN}+B_{MN})$ term and is therefore identical to the GS action on
$AdS_2\times S^2$ \cite{Tseytlin}\cite{Zhu}. 
However, the second line of this action is crucial
for quantization and is absent from the GS action. Using standard
techniques, it was confirmed to one-loop sigma model in
\cite{zw} that the above action is conformally invariant as expected.

\section{Hybrid formalism for compactification to six dimensions}

For compactifications to six dimensions, it is convenient to
choose
$$\t^\a = [ e^{\frac{1}{2}\phi}\Sigma^{+++++},
e^{\frac{1}{2}\phi}\Sigma^{--+++},
e^{\frac{1}{2}\phi}\Sigma^{-+-++},
e^{\frac{1}{2}\phi}\Sigma^{+--++}],$$ 
$$p_\a = [ e^{-\frac{1}{2}\phi}\Sigma^{-----},
e^{-\frac{1}{2}\phi}\Sigma^{++---},
e^{-\frac{1}{2}\phi}\Sigma^{+-+--},
e^{-\frac{1}{2}\phi}\Sigma^{-++--}]$$ 
to be fundamental variables where $\a=1$ to 4.
For compactifications of the superstring which
preserve six-dimensional supersymmetry, the four-dimensional
compactification manifold
can be represented by a $c=6$  N=2 superconformal field theory whose
N=2 generators will be denoted $[T_C, G^+_C, G^-_C, J_C]$.
As before,
for the compactification-dependent worldsheet fields to have no singular OPE's
with $\t^\a$ and $\hat\t^{\dot\a}$, they need to be twisted
by a factor $e^{N\kappa }$ where $\p\kappa=\p\phi + \eta\xi$
and $N$ is the charge of the
field with respect to $J_C$. Furthermore, the 
$c=6$ N=2 generators need to be redefined as
$$[T_C, G^+_C, G^-_C, J_C] \to [T_C +\p^2\kappa-\p\kappa J_C,
e^\kappa G^+_C, e^{-\kappa} G^-_C, J_C +2\p\kappa]$$
which still generates a $c=6$ N=2 algebra.

There are two chiral bosons, $\rho$ and $\sigma$, defined by
$$\p\rho= -2\p\phi -\xi\eta - J_{C}, \quad \p\sigma=i bc $$
which have no singular OPE's with the other fields. Since
$[x^m, \t^\a, p_\a, \rho, \sigma]$
contain the same degrees of freedom as the compactification-independent
RNS variables $[x^m, \psi^m, b, c, \beta, \gamma]$ where $m=0$ to 5,
any vertex operator constructed out of RNS variables can
be rewritten in terms of hybrid variables and vice versa.
After performing a similarity transformation on the hybrid variables,
the twisted N=2 generators described in the previous section are mapped to
$$T =\frac{1}{2} 
\p x^m \p x_m + p_\a \p\theta^\a
+ \frac{1}{2}\p\rho\p\rho 
+ \frac{1}{2}\p\sigma\p\sigma
+ \frac{3}{2}\p^2(\rho +i\sigma) + T_C ,$$
$$G^+ = j_{BRST} = -e^{-2\rho-i\sigma}(p)^4 
+{i\over 2}
e^{-\rho} p_\a p_\b \p x^{\a\b} $$
$$
+e^{i\sigma}( \half\p x^m \p x_m +
p_\a\p \t^\a  +\half\p(\rho+i\sigma)\p(\rho+i\sigma)-\half\p^2(\rho+i\sigma))
+ G^{+}_{C} , $$
$$G^{-}  = b= 
e^{-i\sigma}+ G^{-}_C ,$$
$$J=j_{ghost} =\p(\rho+i\sigma)~+J_{C} ,$$
where
$(p)^4={1\over 24}\e^{\a\b\c\d}p_\a p_\b p_\c p_\d$,
$x^m$ has been written in bispinor
notation as
$x^{\a\b}=(\sigma_m)^{\a\b} x^m$, and $(\s_m)^{\a\b}$
are the six-dimensional Pauli matrices satisfying
$$(\sigma_m)^{\a\b}(\sigma_n)_{\b\c} +
(\sigma_n)^{\a\b}(\sigma_m)_{\b\c} =2\eta_{mn}\d^\a_\g$$
with
$(\s_m)_{\a\b}$ defined as
$(\s_m)_{\a\b}
=\half\epsilon_{\a\b\c\d} (\s_m)^{\c\d}.$

Although physical vertex operators can be defined as U(1)-neutral primary
fields with respect to the above N=2 generators, they would not be manifestly
spacetime-supersymmetric since they depend on 
only half of the usual eight $\t^{\a j}$ variables
of six-dimensional superspace where $j=\pm$. If one calls $\t^{\a -}=\t^\a$
where $\t^\a$ is
defined above, the complex conjugates $\t^{\a +}$ are missing.
To make all six-dimensional supersymmetries manifest, one therefore 
needs to add
$\t^{\a +}$ to the hybrid variables, as well as their conjugate momentum
$p_{\a +}$. Since these variables are not related by a field redefinition
to RNS variables, one needs to simultaneously introduce new fermionic
gauge invariances 
which allow $\t^{\a +}$ and $p_{\a +}$ to be gauged away \cite{six}.
These new variables will be defined to have no singular OPE's with
the other fields and to satisfy $\t^{\a +}(y) p_{\b +}(z)\to (y-z)^{-1}
\delta^\a_\b$.

The fermionic gauge invariances will be generated by
the first-class constraints 
$$D_\a = d_{\a +} - e^{\-\rho-i\sigma} d_{\a -} =0$$
where $d_{\a -}=p_\a$ and $d_{\a +} = 
p_{\a +}-{i} \t^{\b -} \p x_{\a\b} + \frac{1}{8}\epsilon_{\a\b\gamma\delta}
\t^{\b -}\t^{\gamma -}\p\t^{\delta +}$.
Since $\{D_\a, \t^{\b +}\}=\delta_\a^\b$, this gauge invariance can
be used to fix $\t^{\a +}=0$ and the constraint then fixes
$p_{\a +}= e^{-\rho-i\sigma} p_{\a -} +{i} \t^{\b -} \p x_{\a\b} .$
Furthermore,
the N=2 constraints can be modified to commute with
the $D_\a$ constraints by defining
$$T =\frac{1}{2} 
\p x^m \p x_m + p_{\a j}\p\theta^{\a j} + \p\t^{\a +} D_\a
+ \frac{1}{2}\p\rho\p\rho 
+ \frac{1}{2}\p\sigma\p\sigma
+ \frac{3}{2}\p^2(\rho +\sigma) + T_C ,$$
$$G^+ = -\frac{1}{24}\epsilon^{\a\b\gamma\delta} 
D_\a (D_\b (D_\c (D_\d (~e^{2\rho+3i\s}~)))) ~+G_C^+,$$
$$G^{-}  = 
e^{-i\sigma}+ G^{-}_C ,$$
$$J =\p(\rho+i\sigma)~+J_{C} ,$$
where $D_\a(Y)$ denotes the contour integral of $D_\a$ around $Y$.
One can check that these N=2 constraints agree with the ones
defined earlier in the gauge $\t^{\a +}=0$.
These N=2 generators, 
together with the $D_\a$ constraints,
provide first-class constraints for $d_{\a j}$  that
can replace the second-class constraints of the six-dimensional GS
superstring.

For the open superstring, the massless compactification-independent
states are described by the gluon and gluino of six-dimensional
super-Yang-Mills.
As shown in \cite{four}, the integrated vertex operator for these states
is given by
$$V = \int dz [ \p Y^M ~A_M + d_{\a  j}~W^{\a j}] 
$$
exactly as predicted by Siegel where $Y^M = (x^m,\t^{\a j})$,
$W^{\a j}$ is the superspace field-strength,
and $A_M$ are the superspace gauge fields. 
In Wess-Zumino gauge when $\t^{\a +}=0$, this
vertex operator is mapped to the standard Neveu-Schwarz and Ramond
vertex operators for the gluon and gluino where the gluino
vertex operator with polarization $W^{\a -}$
is in the $-\frac{1}{2}$ picture, the gluon vertex operator is in the
0 picture, and the gluino vertex operator with polarization $W^{\a +}$ 
is in
the $+\frac{1}{2}$ picture \cite{six}.

Since the closed superstring vertex operator is the holomorphic square
of the open superstring vertex operator, one obtains the following integrated
vertex operator for the massless compactification-independent states
of the Type II superstring:
$$S = \int d^2 z  [ \p Y^M \bar\p Y^N ~(G_{MN} +B_{MN})
 + d_{\a j}\bar\p Y^N E^{\a j}_N + 
\p Y^M \bar d_{\bar \a j} E_M^{\bar \a j} + 
d_{\a j} d_{\bar \b k} F^{\a j\bar \b k}]$$
where $Y^M=
(x^m,\t^{\a j}, \bar\t^{\bar\a j})$,
$\bar\t^{\bar\a j}$ 
is the right-moving analog of the left-moving
$\t^{\a j}$, 
and the lowest components of the superfields $G_{mn}, B_{mn},
E_m^{\a j}, E_m^{\bar \a j}, F^{\a j\bar \b k}$ are the graviton, 
the anti-symmetric
tensor, the two gravitini, and the Ramond-Ramond bispinor field strengths.

To get the superstring action in a curved background, one again
interprets the superfields [$G_{MN},B_{MN},E_M^{\a j}, E_M^{\bar\a j}
,F^{ \a j\bar \b k}$]
appearing in $S$ as background superfields. This action is manifestly
super-reparameterization invariant and is expected to
be N=2 worldsheet superconformally invariant when
the background fields satisfy the appropriate equations of motion.
The first term in $S$ is the usual GS action in a curved
background, but the other terms in $S$ proportional to
$d_{\a j}$ and $\bar d_{\bar \a j}$ are required for quantization.
The complete action also contains a contribution from the
compactification-dependent fields, from the chiral bosons,
and from a Fradkin-Tseytlin term constructed in a manner
similar to that of the four-dimensional action \cite{six}.

To obtain the action for the superstring in an $AdS_3\times S^3$
background 
with constant Ramond-Ramond flux $F^{\a j\bar \b k} =  \epsilon^{jk}
\delta^{\a\bar \b}$, there are two approaches. The first approach \cite{six}
resembles the four-dimensional case where 
one simply plugs the appropriate values for the background fields into the
action. After integrating out $d_{\a j}$ and $\bar d_{\bar \a j}$, one
obtains the action
$$
S_{AdS} = \int d^2 z 
[ \eta_{cd} J_z^c J^d_{\bar z} + \frac{1}{2} \epsilon_{jk}
\delta_{ \a j\bar \a k}
(J_z^{\a j} J_{\bar z}^{\bar \a k} - J_{\bar z}^{\a j} J_z^{\bar \a k}) 
$$
$$
+ \epsilon_{jk}\delta_{ \a j\bar \a k}(J_z^{\a j} J_{\bar z}^{\bar \a k} + 
J_{\bar z}^{\a j} J_z^{\bar \a k})]
$$
where $(J^{\a j}, J^{\bar \a k})$ and $J^c$ are the sixteen fermionic currents 
and six bosonic currents $(g^{-1} d g)^A$
constructed from a coset supergroup $g$ taking values in $PSU(1,1|2)
\times PSU(2|2)/SU(2)\times
SU(2)$. The first line of this action comes from the $\p Y^M \bar\p Y^N 
(G_{MN}+B_{MN})$ term and is therefore identical to the GS action on
$AdS_3\times S^3$ \cite{Tseytlin}\cite{rajaraman}. 
However, the second line of this action is crucial
for quantization and is absent from the GS action. Using standard
techniques, it was confirmed to one-loop sigma model in
\cite{zw} that the above action is conformally invariant as expected.

A second approach for constructing the action on $AdS_3\times S^3$ is
to first use the $D_\a$ and $\bar D_{\bar \a}$
constraints to solve for $d^{\a +}$ and $\bar
d^{\bar\a -}$ and to gauge-fix half of the fermionic parameters of the
coset supergroup. 
After integrating out the remaining $d^{\a -}$ and
$\bar d^{\bar\a +}$ variables, one obtains the action \cite{witten}
$$
S_{AdS} = \int d^2 z 
[ \eta_{cd} J_z^c J^d_{\bar z} -(1- e^{-\rho-i\sigma+\bar\rho
+i\bar\sigma})^{-1} (J_{\bar z}^{\a -}+
e^{-\rho-i\sigma} J_{\bar z}^{\a +})
(J_z^{\a +} + e^{\bar\rho+i \bar\sigma} J_{z}^{\a -})]$$
where $(J^{\a -}, J^{\a +})$ and $J^c$ are the eight fermionic 
and six bosonic left-invariant currents $(g^{-1} d g)^A$
constructed from a supergroup $g$ taking values in $PSU(2|2)$.
It was proven to all orders in sigma model loops that
this action is conformally invariant \cite{witten}.

\ack
I would like to thank my collaborators for their contributions, the
organizers of Strings `99 for an enjoyable conference, and CNPq 
grant 300256/94-9 for partial financial support.


\end{document}